\documentclass[twocolumn]{aastex62}

\newcommand{\tm}[1]{{#1 }}
\received{XXXXX X, 2019}
\revised{\today}
\accepted{XXXXXX}
\submitjournal{ApJL}

\shorttitle{Compton-thick AGN with Double-Peaked Narrow Lines}
\shortauthors{Miyaji et al.}

\begin{document}

\title{Torus Constraints in ANEPD-CXO245: A Compton-thick AGN with Double-Peaked Narrow Lines}

\correspondingauthor{Takamitsu Miyaji}
\email{miyaji@astro.unam.mx}

\author[0000-0002-7562-485X]{Takamitsu Miyaji}
\affil{Instituto de Astronom\'ia sede Ensenada \\
 Universidad Nacional Aut\'onoma de M\'exico\\
 Km 107, Carret. Tij.-Ens., Ensenada, 22060, BC, Mexico}
\author{Mart\'in Herrera-Endoqui}
\affil{Instituto de Astronom\'ia sede Ensenada \\
 Universidad Nacional Aut\'onoma de M\'exico\\
 Km 107, Carret. Tij.-Ens., Ensenada, 22060, BC, Mexico}
\author{Mirko Krumpe}
\affil{Leibniz Institut f\"ur Astrophysik, Potsdam \\
  An der Sternwarte 16, 14482 Potsdam, Germany}
\author{Masaki Hanzawa}
\affil{Department of Physics, Kwansei Gakuin University, 2-1 Gakuen, Sanda, Hyogo 669-1337, Japan}
\author{Ayano Shogaki}
\affil{Department of Physics, Kwansei Gakuin University, 2-1 Gakuen, Sanda, Hyogo 669-1337, Japan}
\author[0000-0002-5698-9634]{Shuji Matsuura}
\affil{Department of Physics, Kwansei Gakuin University, 2-1 Gakuen, Sanda, Hyogo 669-1337, Japan}
\author[0000-0002-0114-5581]{Atsushi Tanimoto}
\affil{Department of Astronomy, Kyoto University, Kitashirakawa-Oiwake-cho, Sakyo-ku, Kyoto 606-8502, Japan}
\author[0000-0001-7821-6715]{Yoshihiro Ueda}
\affil{Department of Astronomy, Kyoto University, Kitashirakawa-Oiwake-cho, Sakyo-ku, Kyoto 606-8502, Japan}
\author{Tsuyoshi Ishigaki}
\affil{Department of Science and Technology, Iwate University, 
  3-18-34 Ueda, Morioka, Iwate 020-8550, Japan}
\author{Laia Barrufet} 
\affil{European Space Astronomy Center (ESAC), 28691 Villanueva de la Canada, Spain} 
\author{Hermann Brunner}
\affil{Max-Planck-Institut f\"ur extraterrestrische Physik, 85748 Garching bei M\"unchen, Germany}
\author{Hideo Matsuhara}
\affil{JAXA/ISAS, 3-1-1 Yoshinodai, Sagamihara, Kanagawa, Japan}
\author{Tomotsugu Goto}
\affil{National Tsing Hua University, No. 101, Section 2, Kuang-Fu Road, Hsinchu 30013, Taiwan}
\author{Toshinobu Takagi}
\affil{Japan Space Forum, 3-2-1, Kandasurugadai, 
  Chiyoda-ku, Tokyo 101-0062, Japan}
\author{Chris Pearson} 
\affil{RAL Space, STFC Rutherford Appleton Laboratory, Chilton, Didcot, Oxfordshire OX11 0QX, UK}
\affil{The Open University, Milton Keynes, MK7 6AA, UK}
\author[0000-0002-4193-2539]{Denis Burgarella} 
\affil{Aix-Marseille Université, CNRS, LAM (Laboratoire d'Astrophysique de Marseille) 
  UMR 7326, 13388 Marseille, France}
\author{Nagisa Oi} 
\affil{Tokyo University of Science, 1-3 Kagurazaka, Shinjuku-ku, Tokyo 162-8601, Japan}
\author[0000-0001-6919-1237]{Matthew Malkan} 
\affil{Department of Physics and Astronomy, UCLA, 475 Portola Plaza, Los Angeles, CA 90095-1547, USA}
\author[0000-0002-3531-7863]{Yoshiki Toba}
\affil{Department of Astronomy, Kyoto University, Kitashirakawa-Oiwake-cho, Sakyo-ku, Kyoto 606-8502, Japan}
\affil{Academia Sinica Institute of Astronomy and Astrophysics, 11F of Astronomy-Mathematics Building, AS/NTU, No.1, Section 4, 
  Roosevelt Road, Taipei 10617, Taiwan}
\affil{Research Center for Space and Cosmic Evolution, Ehime University, 2-5 Bunkyo-cho, Matsuyama, Ehime 790-8577, Japan}
\author[0000-0002-7126-691X]{Glenn J. White}
\affil{RAL Space, STFC Rutherford Appleton Laboratory, Chilton, Didcot, Oxfordshire OX11 0QX, UK}
\affil{The Open University, Milton Keynes, MK7 6AA, UK}
\author{Hitoshi Hanami}
\affil{Department of Science and Technology, Iwate University, 
  3-18-34 Ueda, Morioka, Iwate 020-8550, Japan}


\begin{abstract}
 
  We report on the torus constraints of the Compton-thick AGN with double-peaked optical 
  narrow line region (NLR) emission lines, ANEPD-CXO245, at $z=0.449$ in the {\sl AKARI} NEP Deep Field.
  The unique infrared data on this field, including those from the nine-band photometry over 2-24 $\mu$m
  with the {\sl AKARI} Infrared Camera (IRC), and the X-ray spectrum from {\sl Chandra} allow us to
  constrain torus parameters such as the torus optical depth, X-ray absorbing column, torus angular 
  width ($\sigma$) and viewing angle ($i$). We analyze the X-ray spectrum as well as 
  the UV-optical-infrared spectral energy distribution (UOI-SED) with clumpy torus models
  in X-ray \citep[{\sl XCLUMPY};][]{tanimoto19} and infrared \citep[{\sl CLUMPY}; ][]{nenkova08} respectively. 
  From our current data, the constraints on $\sigma$--$i$ 
  from both X-rays and UOI show that the line of sight crosses the torus as expected for a
  type 2 AGN. \tm{We obtain a small X-ray scattering fraction ($<0.1$\%), which suggests narrow torus
    openings, giving preference to the bi-polar outflow picture of the double-peaked profile.}
  Comparing the optical depth of the torus from the UOI-SED and the absorbing column density $N_{\rm H}$ 
  from the X-ray spectrum, we find that the gas-to-dust ratio is $\ga 4$ times larger than the Galactic value.

\end{abstract}




\section{Introduction} \label{sec:intro}
 
 In the course of our multi-wavelength survey on the {\sl AKARI} NEP Deep Field (ANEPD), 
including {\sl Chandra} X-ray observations \citep{K15nep,miyaji17nep},
optical spectroscopy \citep{shogaki_mthesis}, and early UV-optical-infrared (UOI) 
spectral energy distribution (SED) analysis \citep{hanami12}, we have found an optically type 2
Compton-thick (CT) AGN, ANEPD-CXO245 
(hereafter CXO245; $z=0.449$, $[\alpha,\delta]_{\rm J2000}=[17^{\rm h}56^{\rm m}01\fs 69,\,66\degr 35\arcmin 00\farcs 6]$
), which exhibits double-peak optical emission lines from the AGN Narrow-Line 
Region (NLR). 

 About $\sim 1$\% of present-day type 2 AGNs show double-peaked narrow line region (NLR) 
features \citep{liu10}. 
The origin of the double peaked narrow lines can be heterogeneous
and may be caused by dual AGNs, wind-driven outflows, 
radio-jet driven outflows and rotating ring-like NLRs \citep{muellersanchez15}. 
 To discriminate among these scenarios, AGN torus parameters that can be obtained by 
the analysis of the X-ray spectrum and/or UOI-SED can give a clue, in particular, 
to distinguish between the outflow and rotating NLR pictures. In the case of a narrow torus 
opening, it is more difficult for a rotating ring to cross the ionization cone 
and the bi-polar picture would be favorable. If the line of sight is almost perpendicular to the polar axis, 
the two sides of a bi-polar outflow would show similar line-of-sight velocities 
and in this case, the outflow picture would not be favored. In any case, whether 
the bi-polar outflows and/or rotating rings are generally associated with highly absorbed CT-AGNs 
can have implications in their evolution stage. The CT AGNs may be at the stage of starting feedback through outflows
or tidally-disrupted in-falling clouds generating a ring-like structure. 

 Another interesting implication of X-ray spectral and UOI-SED analysis
is the gas-to-dust ratio of the AGN torus, since the torus IR emission is from dust whereas
the X-ray absorption and reflection are produced by gas \citep{ogawa19,tanimoto19}.  
 
 In view of these, we conduct an AGN torus analysis of CXO245 both from our {\sl Chandra} 
X-ray spectrum as well as the UOI SED taking advantage of the unique mid-IR photometric 
bands available in ANEPD. In Sect. \ref{sec:data}, we summarize the dataset used. In
Sect. \ref{sec:anares}, we summarize the key results from the optical emission 
lines and explain our methods and results of individual and joint X-ray 
spectral and UOI SED analyses. Discussions and concluding remarks are 
made in Sect. \ref{sec:disc}. 

 We use $H_0=70\,{\rm km\,s^{-1}\,Mpc^{-1}}$, $\Omega_{\rm m}=0.3$, and $\Omega_{\Lambda}=0.7$ 
throughout this paper.  

\section{Data}
\label{sec:data}
\subsection{UV, Optical and Infrared (UOI) Data}
This object was found as a result of our {\sl AKARI} survey on the North Ecliptic
Pole (NEP) region \citep[AKARI NEP Deep Field; e.g.][]{matsuhara06}, where deep
observations with all the nine bands of the InfraRed Camera (IRC;
$\lambda_{\rm eff}=$ 2,3,4,7,9,11,15,18 \& 24 $\mu$m)
were made. Extensive multi-wavelength images have been obtained on this field by 
ground-based and space-bourne observatories. We use UOI photometric measurements  
from GALEX \citep{burgarella19}, Subaru Telescope Suprime Cam (SCAM) \citep{murata13}, 
Canda-Fracnce-Hawaii Telescope (CFHT) MegaCam and WIRCAM \citep{oi14},  
and Herschel PACS \citep{pearson19}/SPIRE. The SPIRE data, originally 
published by \citet{burgarella19} has been re-analyzed by Pearson et al. (in prep)
and we use the revised photometry. Table \ref{tab:phot} shows a summary of the UOI photometry. 
The optical spectra of CXO245 have been obtained during our KECK (DEIMOS) runs in 2008 and 
2011 and reduced by \citet{shogaki_mthesis} using the {\sf DEIMOS DEEP2 reduction pipeline}.
The spectrum from the 2011 run is shown in Fig. \ref{fig:deimos}.

\begin{deluxetable}{lcccl}
\tabletypesize{\footnotesize}
\tablecaption{UV-Opt.-IR Photometry Used\label{tab:phot}}
\tablecolumns{5}
\tablewidth{0pt}
\tablehead{
 \colhead{Band} & \colhead{$\lambda_{\rm eff}$} & \colhead{Flux} & \colhead{Err.($1\sigma$)} & \colhead{Telescope/Instrument} \\
                & \colhead{${\rm [\mu\,m]}$}   & [mJy] & [mJy] &  
}
\startdata
NUV     & 0.229 & 4.980e-4 & 8.0e-5 &  GALEX        \\                             
u*      & 0.381 & 1.061e-3 & 1.8e-5 &  CFHT/MEGACAM \\        
B       & 0.437 & 2.584e-3 & 8.8e-6 &  SUBARU/SCAM  \\        
V       & 0.545 & 7.553e-3 & 1.6e-5 &  SUBARU/SCAM  \\        
r       & 0.651 & 1.939e-2 & 1.9e-5 &  SUBARU/SCAM  \\        
NB711   & 0.712 & 2.547e-2 & 3.1e-5 &  SUBARU/SCAM  \\        
i       & 0.768 & 3.273e-2 & 2.3e-5 &  SUBARU/SCAM  \\        
z       & 0.919 & 4.526e-2 & 4.6e-5 &  SUBARU/SCAM  \\        
Y       & 1.03  & 7.973e-2 & 5.1e-4 &  CFHT/WIRCAM  \\        
J       & 1.25  & 1.113e-1 & 9.2e-4 &  CFHT/WIRCAM  \\        
Ks      & 2.15  & 1.886e-1 & 1.0e-3 &  CFHT/WIRCAM  \\        
N2      & 2.41  & 2.198e-1 & 3.1e-3 &  AKARI/IRC    \\        
N3      & 3.28  & 1.881e-1 & 2.2e-3 &  AKARI/IRC    \\        
N4      & 4.47  & 1.706e-1 & 2.0e-3 &  AKARI/IRC    \\        
S7      & 7.30  & 4.521e-1 & 1.4e-2 &  AKARI/IRC    \\        
S9W     & 9.22  & 7.238e-1 & 1.8e-2 &  AKARI/IRC    \\        
S11     & 10.9  & 1.036e+0 & 2.3e-2 &  AKARI/IRC    \\        
L15     & 16.2  & 1.562e+0 & 3.7e-2 &  AKARI/IRC    \\        
L18W    & 19.8  & 2.297e+0 & 4.0e-2 &  AKARI/IRC    \\        
L24     & 23.4  & 3.342e+0 & 8.6e-2 &  AKARI/IRC    \\        
PACS100 & 100   & 4.760e+0 & 1.5e+0 &  HERSCHEL/PACS\\        
PACS160 & 160   & 1.768e+1 & 4.5e+0 &  HERSCHEL/PACS\\        
PSW     & 250   & 2.974e+1 & 3.8e+0 &  HERSCHEL/SPIRE\\        
PMW     & 350   & 2.353e+1 & 2.9e+0 &  HERSCHEL/SPIRE\\        
PLW     & 500   & 1.320e+1 & 3.7e+0 &  HERSCHEL/SPIRE\\            
\enddata
\end{deluxetable}

\subsection{X-ray Data and Reduction}

 A major fraction ($\sim 0.25\,{\rm \deg}^2$) of ANEPD has been observed with 
{\sl Chandra} with a total exposure of $\sim 300 {\rm ks}$ \citep{K15nep}. CXO245 is covered 
by the {\sl Chandra} ACIS-I FOVs of seven OBSIDs (see {\it Facilities}; total
exposure $\sim 120$ ks with off-axis angles from 3.3$^\prime$ to 9.6$^\prime$).
The X-ray spectrum of each OBSID has been extracted from a circular region
with a radius corresponding to the larger of 50\% ECF at 3.5 keV (from the Ciao tool \verb|psfsize_srcs|)
or 3.5$^{\prime\prime}$. The background spectrum is extracted from an annulus with inner and outer radii
of 10.5$^{\prime\prime}$ and 55$^{\prime\prime}$ respectively, excluding a $10^{\prime\prime}$ region around
another X-ray source (ANEPD-CXO358). A merged source and a background spectra have been
generated using the {\sf Ciao} tool {\sf combine\_spectrum} with the option \verb|bscale_method=time|.
This option generates both the combined source and background spectra in integer counts per bin 
accompanied by an appropriately weighted mean response matrix and a background scaling factor.
These allow us to fit the background subtracted spectrum with full Poisson statistics (for 
small counts) with the {\sf XSPEC} option \verb|cstat|. In our X-ray spectroscopic analysis,
we use the merged source spectrum with the supporting files created in this step. 
The resulting X-ray spectrum is shown in Fig. \ref{fig:clumpyfit}(a) along with 
the model described in Sect. \ref{sec:clumpyX}.

\section{Analysis and Results}
\label{sec:anares}
\subsection{Optical Emission Lines}
 
\begin{figure}[ht]
\vspace*{0.0cm}
\begin{center}
%
  \includegraphics[width=\columnwidth]{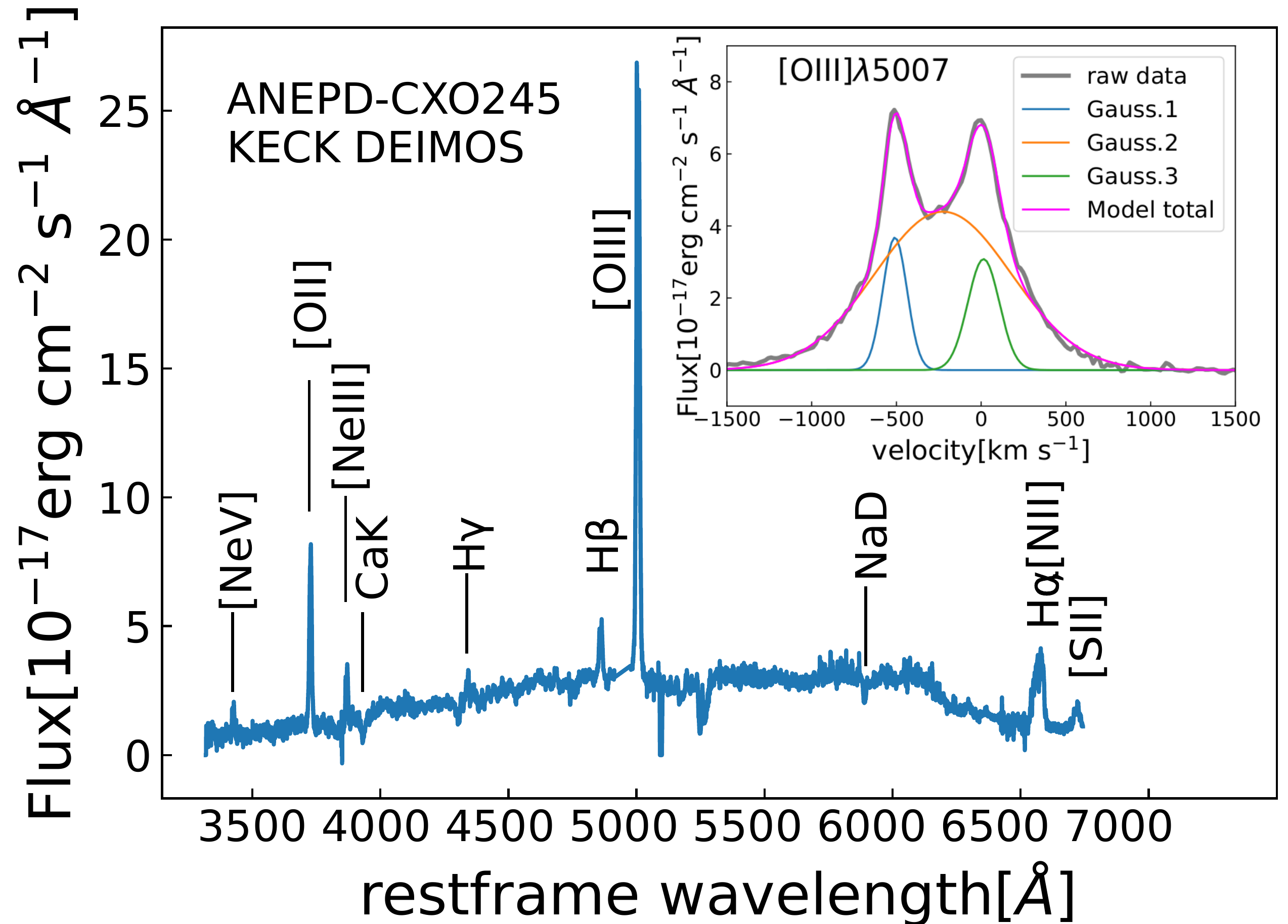}
\end{center}
\caption{
The KECK DEIMOS spectrum of ANEPD-CXO245 from our 2011 in rest frame. 
The inset shows the profile of the double-peaked 
[OIII]$\lambda 5007$ emission line from our 2008 spectrum in the radial velocity space
with a three-Gaussian decomposition model. The raw data, model total, and each model component 
are drawn as labeled.}
\label{fig:deimos}  
\end{figure}

The fluxes of each emission line have been obtained with Gaussian+linear continuum fits.
Multiple Gaussian components are used if needed. The details of the line spectral analysis 
is beyond the scope of this paper. Here we describe the key results of the analysis.
\begin{enumerate}
\item The line ratios of [OIII]$\lambda 5007$/H$\beta \sim 10$ and [NII]$\lambda$ 6583/H$\alpha \sim 1.5$
  are well inside the AGN regime in the BPT diagnostic diagram \citep{BPT}. The spectrum shows 
  [NeV]$\lambda 3425$, which is an an-ambiguous indication of the AGN NLR.  
\item The line profiles of the [OIII]$\lambda 5007$, H$\beta$, and H$\alpha$  
  emission features are all well represented by two narrow
  (FWHM$\sim$150 ${\rm [km\,s^{-1}}]$ each) and a broader (FWHM$\sim$900 ${\rm [km\,s^{-1}]}$) 
  components. The profiles of noisier [NeIII]$\lambda 3869$ and [NeV]$\lambda 3425$ lines
  also show similar double peaks.
  Figure \ref{fig:deimos} (inset) shows the line profile of [OIII]$\lambda 5007$ with
  the best-fit three-Gaussian components as the best example. 
\item The two narrow components are separated by $\sim 500\,{\rm km\,s^{-1}}$ and have similar fluxes.
    The peak of the broad component is just halfway between the two narrower peaks.   
\item The star-formation dominated line [OII]$\lambda 3727$ is single-peaked. Our nominal
redshift ($z=0.499$) is based on this line.
\end{enumerate}

%
%

\subsection{X-ray Spectrum and IR SED: Torus Analysis}

\subsubsection{Clumpy Torus: X-ray Spectrum}
\label{sec:clumpyX} 

 Current popular models of AGN tori involve dusty-gas media
consisting of  ``clumps'' \citep[e.g.][]{elitzur06,nenkova08}.   

 We  first analyze the {\sl Chandra} spectrum of CXO245 using the new 
X-ray Clumpy Torus model {\sf XCLUMPY} \citep{tanimoto19}, which
has the same geometry and geometrical parameters as the {\sf CLUMPY} \citep{nenkova08} model.
Thus direct comparisons with the results of Sect. \ref{sec:clumpyIR} are possible. 
We use the {\sf XSPEC} mode of the form:\\
\verb|  phabs*(zphabs*cabs*cutoffpl+const*cutoffpl| \\
\verb|        +atable{xclumpy_R.fits}|\\
\verb|        +atable{xclumpy_L.fits})|.

The first {\sf phabs} represents the Galactic absorption towards the source direction 
and its column density is fixed to $N_{\rm H,Gal}=4\times 10^{20}{\rm cm^{-2}}$ \citep{nhmap}. 
\tm{The first and second term in the parenthesis are transmitted and scattered primary continuum 
  respectively. The former is subject to a line-of-sight photoelectric absorption ({\sf zphabs}) and
  a Compton scattering ({\sf cabs}) through the torus. The latter expresses that the fraction
  $f_{\rm X,sct}$ (represented by a {\sf const}) is scattered by electrons in thin plasma above and below
  the polar torus openings.}
 The {\sf XSPEC} table models \verb|xclumpy_R.fits| and \verb|xclumpy_L.fits| provide the continuum 
 and the emission line (including fluorescent emission lines from elements up to $Z=30$, dominated by
 Fe K$\alpha$) components of the X-ray reflection from the clumpy torus 
respectively. The normalization and photon index of the primary X-ray continuum are  
free parameters, where the latter is allowed to vary within $1.5\leq \Gamma \leq 2.5$, 
while its cutoff energy is fixed to $E_{\rm c}=300\,{\rm keV}$ \tm{\citep{koss17,ricci18}.
  These parameters are common to the reprocessed, transmitted, and scattered components.}
The solar abundance \citep{anders89} 
is assumed. The redshift parameter of the model components that require one are fixed to $z=0.449$.
Spectral fits are made in channel energies of $0.5-7.0\,{\rm keV}$ using a Markov Chain Monte Carlo
(MCMC) chain with the length of 40,000 (using XSPEC's {\sf chain} command).
 In the current version of {\sf XCLUMPY}, the number of 
clumps along the equator, the ratio of the outer to inner radii, 
and the radial clumpy distribution index are fixed to $N_{\rm clump}^{\rm Equ}=10$, $Y=20$, 
and $q=0.5$ respectively. The parameter ranges covered by the model 
implementation for the equatorial column density, torus width and viewing angle are 
$23\leq \log N_{\rm H}^{\rm Equ}\leq 26$, $10\degr\leq \sigma \leq 70\degr$ and 
$20\degr\leq i \leq 87\degr$ respectively.
   
\begin{figure*}[ht]
\vspace*{0.0cm}
\begin{center}
\hbox{\hspace*{0.0cm}  
  \includegraphics[height=0.345\textwidth]{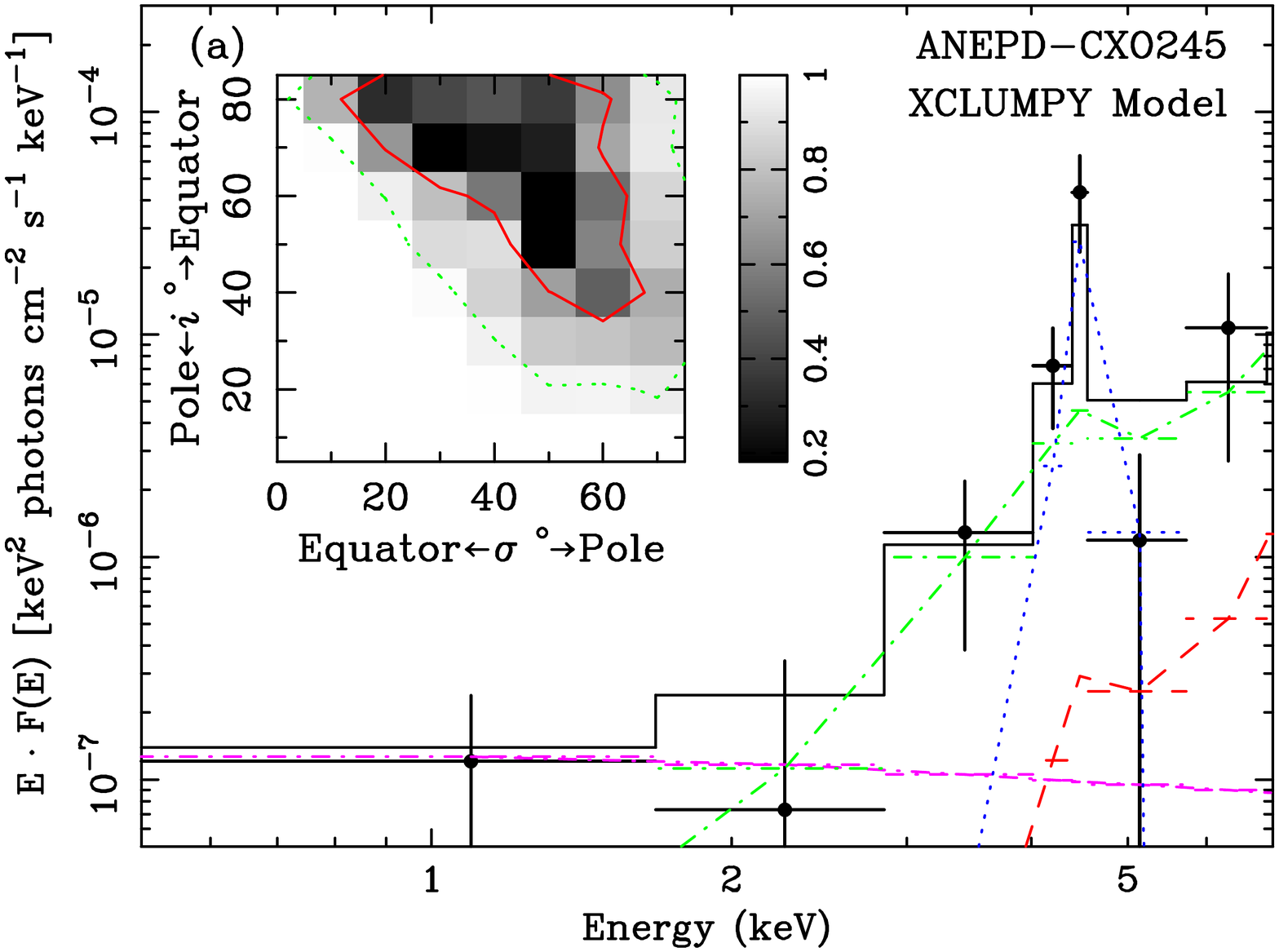}
  \includegraphics[height=0.345\textwidth]{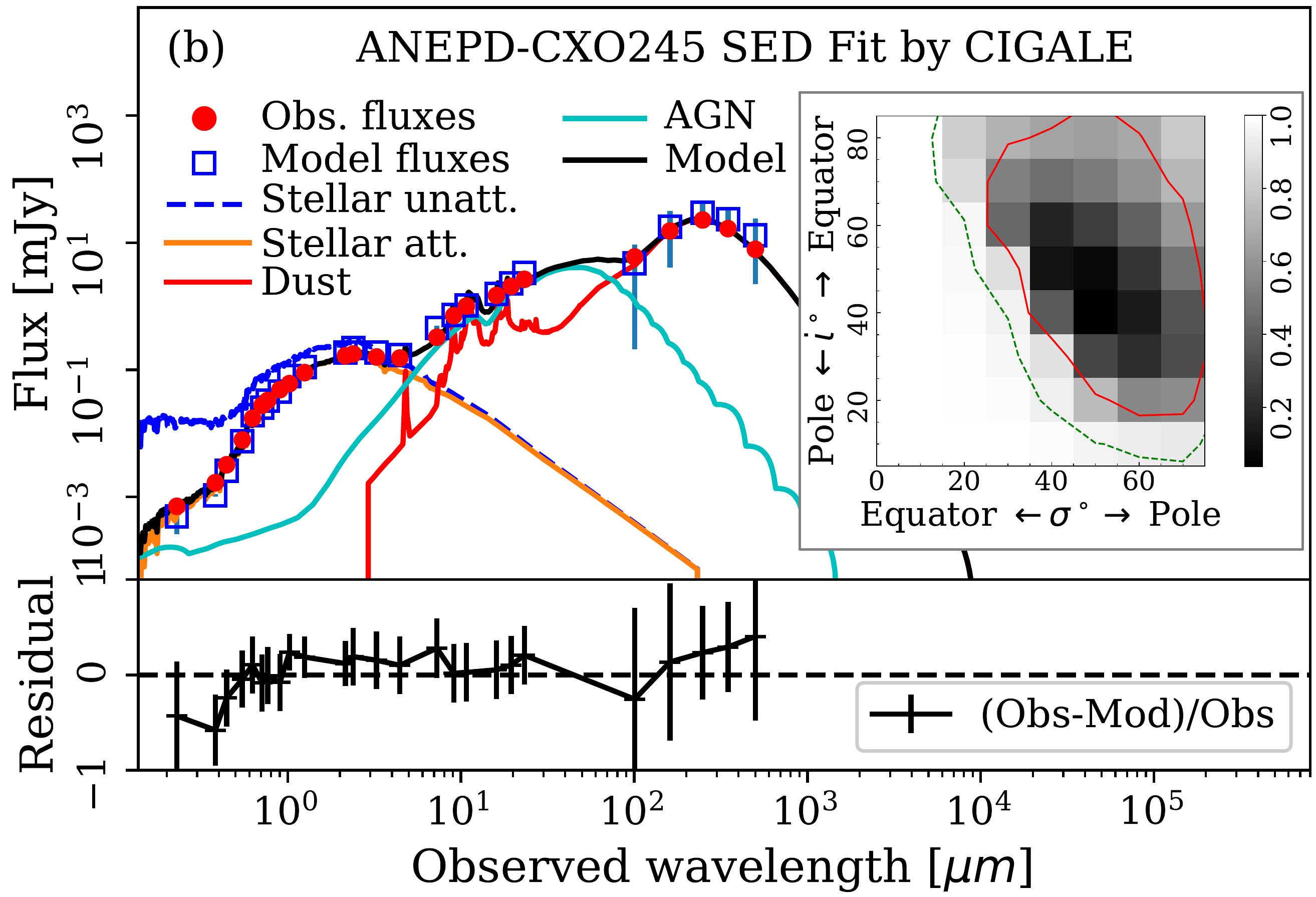}
}\vspace{-1.5\baselineskip}
\end{center}
\caption{
(a) The unfolded $E\cdot F(E)$ representation of our {\sl Chandra} ACIS-I 
spectrum of ANEPD-CXO245. The black filled circles with error bars show the 
observed data binned to at least 2$\sigma$/bin up to 80 ACIS PI channels. 
The binning is for display only. The solid black histogram shows the best-fit model 
described in \ref{sec:clumpyX}. Also plotted are the reflected continuum (green dot-dashed line), 
reflected \tm{fluorescence line} (blue dotted line), \tm{transmitted} (red dashed line)
and \tm{scattered} (magenta dot-dot-dashed line) components. 
The inserted box shows the integrated 
probability image and confidence contours in the $\sigma$ -- $i$ space in
grids of $10\times 10\,{\rm deg^2/pixel}$. The contours correspond to integrated marginal 
probabilities of 68\% (red solid contour) and 95\% (green dotted contour). 
(b) The Optical-IR data fitted with {\sf CIGALE} 
with the {\sf CLUMPY} implementation. The photometry data and best-fit model
with the contribution of each component are shown as labeled with residuals.
The curve labeled ``Dust'' refers to the dust emission from star-formation
activity, while ``AGN'' refers to the AGN torus dust emission from {\sl CLUMPY}. 
The inset shows the integral probability image and the confidence contours at 
the same levels as in panel (a).
}  
\label{fig:clumpyfit}
\end{figure*}

Table \ref{tab:models} shows the best-fit parameters and the 90\% confidence ranges 
obtained from the MCMC chain.  Figure~\ref{fig:clumpyfit}(a) shows the best-fit model 
and the contribution of various components with the unfolded ACIS spectrum. 
 Figure~\ref{fig:clumpyfit}(a) (inset) shows the integrated probability grayscale image and 
 its contours (see caption) in the $\sigma$-$i$ space. Because the available solid angle per viewing angle
 ($i$) is proportional to $\sin i$, we
use $\sin i$ as a prior. Practically, we calculate the 90\% ranges 
from the chain points weighted by the prior. Likewise, the marginal probability in each bin 
$P_{\rm X}(\sigma_j,i_k)$ by accumulating the weighted chain points and normalizing.  
The  integrated probability $I_{\rm X} (\sigma_j,i_k)$ is obtained by iterating, in the order of 
decreasing $P_{\rm X}(\sigma_j,i_k)$:   
\begin{equation}
I_{\rm X}(\sigma_j,i_k)=P_{\rm X}(\sigma_{j},i_{k})+I_{\rm X,prev}, 
\label{eq:intprob}
\end{equation}
where $I_{\rm X,prev}$ is the integrated probability from the previous step
(or 0 in the first step).

The spectrum shows a strong Fe K$\alpha$ line characteristic of a Compton-thick (CT) 
AGN. The derived column densities (both equatorial and line-of sight) correspond
to $N_{\rm H}>10^{24} {\rm cm^{-2}}$ and thus CXO245 can be classified as a CT-AGN.
The confidence contours of Fig. \ref{fig:clumpyfit} and Table \ref{tab:models} 
show that the line-of-sight viewing angle cannot be too close to the pole
($i>30\degr$; 90\% lower limit).   

\begin{deluxetable}{cccc}
\tabletypesize{\footnotesize}
\tablecaption{Model Parameters\tablenotemark{a}\label{tab:models}}
\tablecolumns{4}
\tablewidth{0pt}
\tablehead{
 \colhead{Param.} & \colhead{X-ray Spectrum} & \colhead{UOI SED} & \colhead{Joint} 
}
\startdata
$\log N_{\rm H}^{\rm Equ}$ \tablenotemark{b} & 24.7 (24.5;25.9*) & $\dots$ & $\dots$ \\
$\tau_{\rm V}N_0$ \tablenotemark{c}    & $\dots$ & 400(400;400) & $\dots$  \\
$\sigma$ \tablenotemark{d} &  55 (18;69*) & 50 (20*;70*) & 50 (20*;70*) \\
$i$ \tablenotemark{e}  & 49 (30;85*) & 40 (20;90*) &  50 (30;80*)\\
$\Gamma$ \tablenotemark{f} & 2.2 (1.5;2.4) & $\dots$ & $\dots$ \\
$\log f_{\rm X,sct}$ \tablenotemark{g}  &  -4.0 (-6.0*;-3.0) & $\dots$ & $\dots$ \\
$f_{X,15}$ \tablenotemark{h}      &  8  (5;9) & $\dots$ & $\dots$ \\
$\log L_{\rm X}^{\rm int}$ \tablenotemark{i}  & 44.7 (44.4;45.7) & $\dots$ & $\dots$  \\
$\log L_{\rm AGN}^{\rm IR}$ \tablenotemark{j}  &  $\ldots$ & 44.6 (44.5;44.8) & $\dots$\\
$f_{\rm AGN}^{\rm IR}$ \tablenotemark{k}  & $\dots$  & 0.5 (0.5;0.6)  &  $\dots$ \\
\enddata
\tablenotetext{a}{Best fit values with 90\% confidence range in one parameter in the parentheses.
The bounds that are pegged at model limits are marked with an '*'.}
\tablenotetext{b}{Torus column density ${\rm cm^{-2}}$ at the equator.} 
\tablenotetext{c}{Total optical depth of clumps through the equator at $\lambda=0.55 {\rm \mu m}$.}
\tablenotetext{d}{Torus angular width in degrees.}
\tablenotetext{e}{Viewing angle from the pole in degrees.}
\tablenotetext{f}{Photon index of the primary X-ray continuum.}
\tablenotetext{g}{\tm{X-ray scattering fraction}}
\tablenotetext{h}{X-ray (0.5-7 keV) flux in $10^{-15}\,{\rm erg\,s^{-1}\,cm^{-2}}$ from the best-fit model.}
\tablenotetext{i}{Intrinsic rest frame 2-10 keV luminosity in $\,{\rm erg\,s^{-1}}$ of the primary X-ray continuum.}
\tablenotetext{j}{Infrared luminosity in ${\rm erg\,s^{-1}}$ from the AGN torus.}
\tablenotetext{k}{$f_{\rm AGN}=L_{\rm IR}^{\rm AGN}/(L_{\rm IR}^{\rm SF}+L_{\rm IR}^{\rm AGN})$, where $L_{\rm IR}^{\rm SF}$ is 
  the dust IR luminosity from star formation.}
\end{deluxetable}

\subsubsection{Clumpy Torus: UV-Optical-Infrared (UOI) SED}
\label{sec:clumpyIR} 

We also investigate the AGN torus constraints from the UOI-SED 
($\approx 0.2-1000\,{\rm \mu m}$) of CXO245 in the framework 
of the clumpy torus model {\sf CLUMPY} \citep{nenkova08}. For this purpose, we have made a 
modification to the {\sf CIGALE} package \citep{noll09,boquien19} to include an 
implementation of {\sf CLUMPY}. To make the consistent analysis with the {\sf XCLUMPY} 
X-ray spectrum, we search for best fit parameters assuming $N_0=10$, $Y=20$, 
and $q=0.5$. In the SED fit, we use the  
galaxy stellar component \citep{bruzual03} with a \citet{salpeterIMF} IMF,  
double exponentially-decaying star-formation history and an attenuation by \citep{charlot00}.
For the dust emission models, we use the \citet{dale14} model for the star-formation   
and {\sf CLUMPY} for the AGN torus. The optical part is included in the fits, because the 
star-formation dust component in the IR and the dust attenuation of the star light are energetically 
connected. This helps make a better separation of the AGN and star formation IR components.

 There are certain limitations in the best-fit and parameter error search in {\sf CIGALE}.
For table models, {\sf CIGALE} only allows us to evaluate $\chi^2$ at the grid points in the table 
and no interpolations are made, unlike the X-ray spectral analysis using {\sf XSPEC}. 
The MCMC is not implemented either. Thus the best fit values and bounds 
are among these grid points. In our implementation, the grids of the free geometrical 
parameters are $\sigma=20\degr$--$70\degr$ and $i=0\degr$--$90\degr$ in every 10$\degr$ 
respectively. A common approach in determining a 90\% confidence error range
is to use the $\Delta \chi^2 <2.7$ criterion. However, especially for $\sigma$ and $i$,
parameters are often pegged at the model limits and therefore this criterion does not 
properly indicate the true 90\% probability range.  Thus, we determine the  
90\% confidence range $(p-; p+)$ of the parameter $p$ by $C(<p-)\sim 0.05$ and
$C(<p+)\sim 0.95$ respectively, where $C$ is the cumulative probability:
\begin{equation}
C(<p)=\frac{\sum_{p_k\leq p}\sin i_k\exp\left[-\chi^2(p_k)/2\right]}{\sum_{{\rm all\,} k}\sin i_k\exp\left[-\chi^2(p_k)/2\right]}.  
\label{eq:Pcum}
\end{equation}
Due to computational limitations, we take $\chi^2(p_i)$ as the minimum 
value at $p=p_i$ where all other parameters are allowed to vary, rather than 
the marginal probability, and $i_k$ is the best-fit viewing angle when $p$ is
fixed to $p_k$. The 90\% confidence ranges are approximate because of the 
discreteness of the parameter grid. 
   
Likewise, the probability at each point of the two-parameter space  $(\sigma_j,i_k)$ 
is determined by:
\begin{equation}
P_{\rm UOI}(\sigma_j,i_k)=\frac{\sin i_k \exp\left[-\chi^2(\sigma_j,i_k)/2\right]}
{\sum_{\rm j^\prime k^\prime} \sin i_{k^\prime} \exp\left[-\chi^2(\sigma_{j^\prime},i_{k^\prime})/2\right]},  
\end{equation}
where $\chi^2(\sigma_j,i_k)$ is the minimum $\chi^2$ value at $(\sigma,i)=(\sigma_j,i_k)$ with respect to all 
other parameters. The sum in the denominator is for all the grid points in ($\sigma,i$). Then the 
integrated probability $I_{\rm UOI} (\sigma_j,i_k)$ is obtained in the same manner as Eq. \ref{eq:intprob}.
The resulting best-fit parameters and the 90\% confidence ranges in one parameter for the AGN torus are shown 
in Table \ref{tab:models}. 
Figure~\ref{fig:clumpyfit}(b)(inset)  shows the integrated probability grayscale 
image in the grids mentioned above with contours.

\subsubsection{X-ray -- UOI Joint Torus Constraints}
\label{sec:joint}
\begin{figure}[ht]
\vspace*{0.0cm}
\begin{center}
   \includegraphics[width=0.9\columnwidth]{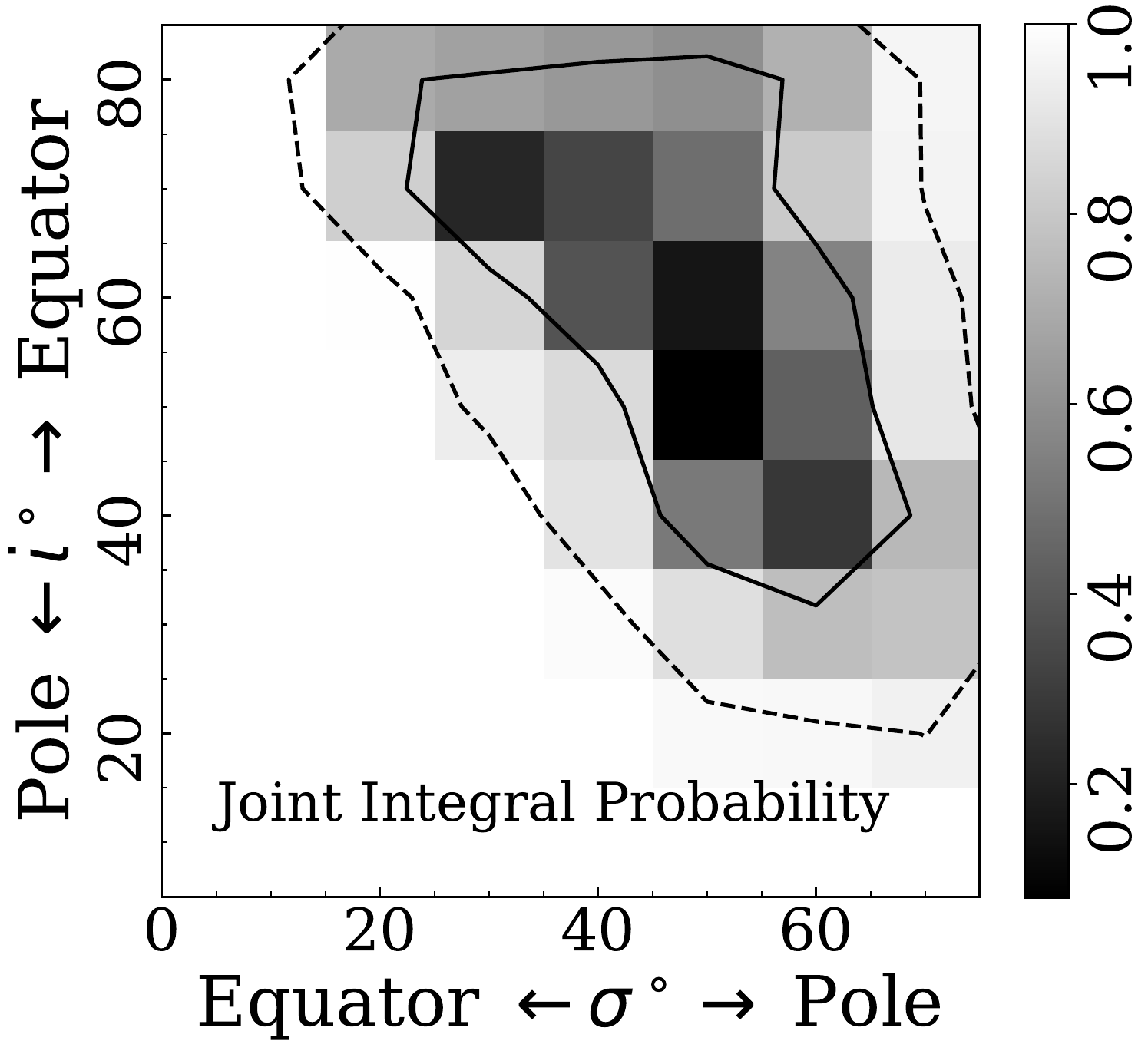}
\end{center}
\caption{
 The integrated probability in the $\sigma$--$i$ space from the joint X-ray and UOI
analysis. The solid and dashed contours represent 68\% and 95\% ranges for the two
parameters respectively.
}  
\label{fig:joint}
\end{figure}

The X-ray spectrum and UOI SED give independent probes
of the torus parameters. The {\sl AKARI} IRC and {\sl Chandra} observations 
were made during 2006 and 2010-2011 respectively and we do not expect 
significant changes in the torus properties between these observations.
Thus we also explore the joint constraints of the torus parameters. 
In the current implementation, the parameters that are common to both 
{\sf XCLUMPY} and {\sf CLUMPY} are $\sigma$ and $i$. 
The joint probability map is calculated by
\begin{equation}
P_{\rm Joint}(\sigma_j,i_k)=
\frac{P_{\rm X}(\sigma_j,i_k)P_{\rm UOI}(\sigma_j,i_k)}
{\sum_{j^\prime k^\prime}P_{\rm X}(\sigma_{j^\prime},i_{k^\prime})P_{\rm UOI}(\sigma_{j^\prime},i_{k^\prime})},
\end{equation}
where the sum is over all pixels in the $(\sigma,i)$ space.
The integrated joint probability image $I_{\rm Joint}$, calculated from $P_{\rm Joint}$ 
in the same manner as Eq. \ref{eq:intprob}, and is shown in 
Fig. \ref{fig:joint}. 

 We note that the new results by Tanimoto et al. (in prep) on the X-ray and IR clumpy 
torus analyses of a sample of 10 nearby type 2 AGNs show inconsistencies of 
$\sigma$ and $i$ values between those obtained by X-ray and IR in some objects. 
Thus the results of the joint constraints should  be used with caution.    

\section{Discussion and Concluding Remark}
\label{sec:disc}

 The developments of modern AGN torus models, both in the infrared and X-rays, have 
opened up the possibility of constraining its geometric parameters such as the torus
angular width and the viewing angle, in addition to the optical depth (UOI) and 
the X-ray absorption column density.  
 
 In our UOI dataset, $\log L_{\rm AGN}^{\rm IR}$ and $\tau_{\rm V}N_0$ 
are well constrained. We verify that $\log L_{\rm AGN}^{\rm IR}$ changes 
very little when we use other models of torus and the star formation dust components 
\citep{fritz06,schreiber16}. With {\sf CLUMPY}, we find $\tau_{\rm V}N_0 = 400$ as the best fit 
among the model's grid points and the neighboring grid values of $200$ and $600$ 
are strongly excluded.  In the X-ray analysis, we obtain 
$N_{\rm H}^{\rm Equ}\ga 4\times 10^{24}{\rm cm^{-2}}$, where the upper bound 
is unconstrained. Thus we obtain 
$N_{\rm H}^{\rm Equ}/A_{\rm V}^{Eq}\ga 3\times 10^{21}{\rm cm^{-2}\,mag^{-1}}$ 
($A_{\rm V}=2.5\tau_{\rm V}N_{\rm 0}/\ln(10)$). The comparison of this ratio with the Galactic value  
\citep[$N_{\rm H}/A_{\rm V}=1.87\times 10^{21} {\rm cm^{-2}\,mag^{-1}}$;][]{draine03},
implies that the gas-to-dust ratio of the CXO245 torus is at least 
4 times larger than that of the Galaxy. This is consistent with 
the results from some other works. \citet{tanimoto19} 
has found a gas-to-dust ratio of $\sim 26$ times the Galactic value
for the nearby CT-AGN the Circinus galaxy. 
New results from a systematic study of 10 additional nearby Seyfert 2 
galaxies with {\sf XCLUMPY} and {\sl CLUMPY} (Tanimoto et al. in prep) include 
measurements of two other CT-AGNs, one of which shows a larger 
$N_{\rm H}/A_{\rm V}$ value than the Galactic one. The comparison of the silicate 
absorption depth $\tau_{9.7}$ at $9.7\,\mu{\rm m}$ and $N_{\rm H}$ by 
\citet{gonzalezmartin13} shows systematically higher $N_{\rm H}$
than expected from $\tau_{9.7}$ expected from the Galatctic
gas-to-dust ratio for obscured AGNs.  

 The constraints on $\sigma$ and $i$ are much looser. There are, however,
some meaningful constraints. The X-ray analysis strongly excludes
$90\degr-i\ga \sigma$, meaning that the line of sight crosses the torus 
material, as expected for type 2 AGNs. 
We also obtain a lower limit to the viewing angle ($i>30\degr$), 
excluding a line of sight that is close to the polar axis. 
The UOI-SED analysis shows a similar trend. 

One of our original motivations of this work was to obtain constraints of these angles to give clues to 
discriminate between the bi-polar outflow and a rotating ring origins of the double-peaked 
NLR lines (Sect.~\ref{sec:intro}). \tm{The constraints of $i$, and $\sigma$ themselves,
  neither X-ray spectrum nor UOI-SED can suggest which picture is preferred.
On the other hand, the very small scattering fraction ($f_{\rm X}\la 0.1\%$) from
our X-ray spectral analysis, suggests a small opening angle (large $\sigma$). While
$f_{\rm X}$--$\sigma$ relation has not yet been calibrated\citet{ueda07,yamada19},
the rather small scattering fraction suggests some preference to the bi-polar outflow picture.}

      

 The 9-band photometric data with {\sl AKARI} IRC available in the {\sl AKARI} NEP Deep and Wide 
fields have made torus analyses with the UOI SED fit possible for CT AGNs across a wide redshift 
range. These can then be compared and/or combined with the X-ray torus analysis, as demonstrated
in this paper. By the analyses on both sides, we obtain a constraint on the gas-to-dust ratio of the AGN 
torus and loose constraints on the torus width and viewing angles. We are planning to extend this work to the 
$\sim 5.4\, {\rm deg}^2$ {\sl AKARI} NEP Wide Field by combining the {\sl AKARI} IRC and supporting UOI data
and the scheduled deep exposures with the recently launched {\sl eROSITA/ART-XC} \citep{erositaSB,artxc18}
in the NEP region. That would provide the candidates for further {\sl Chandra}, {\sl XMM-Newton}  
and {\sl JWST}, and, on a longer timescale, {\sl Athena} observations.


\acknowledgments

The scientific results reported in this article are based on observations made by the {\sl Chandra} 
X-ray Observatory, {\sl AKARI}, the {\sl Herschel} Space Observatory, the Galaxy Evolution Explorer (GALEX),  
the Subaru Telescope, and W.M. Keck Observatory. TM and MHE are
supported by CONACyT 252531 and UNAM-DGAPA PAPIIT IN111319. MK acknowledges support from DLR grant 
50OR1904. GJW gratefully acknowledges support of an Emeritus Fellowship from The Leverhulme Trust.
SM thanks M. Kusunose for helpful discussions on spectral analysis.

%

\vspace{5mm}
\facilities{Chandra (ACIS-I: 10443, 11999, 12931, 12932, 12934, 12935 \& 13244), 
AKARI (IRC), Subaru (Suprime Cam), Keck (DEIMOS),
Herschel (PACS,SPIRE), GALEX}


\software{
          Ciao 4.11\footnote{\url{http://cxc.harvaed.edu/ciao}}, 
          HEASOFT 6.25 (incl.  XSPEC 12.0.1)\footnote{\url{http://https://heasarc.gsfc.nasa.gov/docs/software.html}},
          CIGALE 2018.0\footnote{\url{https://cigale.lam.fr/}}, DEIMOS DEEP2 reduction pipeline
          \footnote{\url{https://www2.keck.hawaii.edu/inst/deimos/pipeline.html}}          
          }

\end{document}